# Non uniqueness of parameters extracted from resonant second-order nonlinear optical spectroscopies

*Bertrand Busson\* and Abderrahmane Tadjeddine*

Laboratoire de Chimie Physique; CNRS, Université Paris-Sud 11; Bâtiment 201 Porte 2; 91405 Orsay; France

Email address: Bertrand.Busson@u-psud.fr; Phone: +33 1 69 15 32 75; Fax: +33 1 69 15 33 28

Abstract

Experimental data from second-order nonlinear optical spectroscopies (SFG, DFG, SHG) provide parameters relevant to the physical chemistry of interfaces and thin films. We show that there are in general $2^N$ or $2^{N-1}$ equivalent sets of parameters to fit an experimental curve comprising N resonant features, of vibrational or electronic origin for example. We provide the algorithm to calculate these sets, among which the most appropriate has to be selected. The main consequences deal with the existence of "ghost resonances", the need of a critical analysis of fit results and the procedure to search for better sets of parameters coherent with applied constraints.

For the past years, infrared-visible Sum-Frequency Generation (SFG) has become a widely used investigation technique for the chemical analysis of interfaces and surfaces. A lot of technical work has been done to improve the experimental set-ups dedicated to the production and detection of SFG photons.

In the nonlinear SFG process, two monochromatic light sources (frequencies $\omega_1$ and $\omega_2$) create a second-order polarization when interacting with a material. At the lowest-order (i.e. dipolar) approximation, this polarization is proportional to the product of the incoming electric fields through the second-order susceptibility tensor $\chi^{(2)}$, which accounts for the material's properties. The macroscopic polarization and susceptibility account for spatial averages of microscopic second-order dipole moments and hyperpolarisabilities ($\beta$), respectively. The second-order polarization may oscillate at the sum ($\omega_1+\omega_2$, SFG) or the difference ($|\omega_1-\omega_2|$, DFG) of the incoming frequencies, acting as a coherent source of new beams at these frequencies. As such processes have a low cross section, they require the beams to have a high energy density, implying the use of short pulsed lasers. Infrared-visible SFG/DFG spectroscopy uses one beam in the visible range, usually with fixed frequency, which makes the detection of the few SFG photons easier, and a tunable (or broadband) one in the infrared (IR) frequency range. The SFG/DFG process becomes resonant (and therefore amplified) when the energy of the infrared beam matches that of an IR and Raman active vibrational transition of the material, making it a vibrational spectroscopy. The main advantage of the nonlinear vibrational spectroscopies as compared to the linear ones (e.g. IR absorption, Raman spectroscopy) lies in a fundamental property of second-order nonlinear optical (NLO) processes, which vanish within media possessing an inversion symmetry. The direct consequence is that SFG/DFG in centrosymmetric materials is only produced where the symmetry is broken, i.e. at the interface between two media. SFG/DFG spectroscopy is therefore intrinsically specific of interfaces. This has lead to the continuous spreading of this spectroscopic tool over time and its application to various kinds of interfaces — molecular monolayers adsorbed at the surface of liquids[1], solids[2] and nanoparticles[3-6]; the surface of liquids[7]; thin molecular films[8] — under diverse environments, for example in vacuum[9], in catalytic conditions[10] or under electrochemical control[11]. Additional refinements include SFG imaging[12-14], two-color and doubly resonant SFG[2,15,16], and SFG near-field microscopy[17,18].

Although IR-visible SFG is mostly used as a vibrational spectroscopy, resonances may occur with any of the three energies involved (i.e. IR, visible, SFG). Resonant second-harmonic generation (SHG) may also be handled within the same theoretical frame, as a special case of SFG.

The theoretical background for any resonant and nonresonant SFG/DFG process has been established in details in many books[19,20]. The intensity of the generated beam is described by

$$I(\omega_G) \propto \left|\chi^{(2)}\right|^2 I(\omega_1) I(\omega_2) \qquad (1)$$

where subscripts 1 and 2 account for the two incoming beams (degenerate in the SHG case), and $\omega_G$ stands for either $\omega_1+\omega_2$, $|\omega_1-\omega_2|$, $2\omega_1$ or $2\omega_2$. Depending on the polarizations of the three beams and the symmetries of the interface, the susceptibility term $\chi^{(2)}$ comprises in general several tensor components.

Resonant processes are embedded into $\chi^{(2)}$ and take the form of Lorentzian functions of the resonant frequencies[21,22]. The Lorentzian description has been widely adopted as a consequence of the quantum perturbation calculation of the nonlinear susceptibility[20], even if it does not account for the inhomogeneous broadening inside the sample. Doubly resonant processes give birth to products of such Lorentzian contributions[22,23].

In a general way, the resonant susceptibility is

$$\chi_R^{(2)} = \sum_{i=1}^{N} \frac{A_i}{\omega - \Omega_i \pm i\Gamma_i} \qquad (2)$$

where $\omega$ is the tunable experimental frequency involved in the resonance (i.e. $\omega_G$, $\omega_1$ or $\omega_2$). Resonances are described by their frequencies $\Omega_i$, damping constants $\Gamma_i$ and amplitudes $A_i$. Coefficients $A_i$ are in general complex quantities:

$$A_i = \tilde{A}_i e^{i\Phi_i} \qquad (3)$$

The ± sign evidences the difference between SFG/SHG (+) and DFG (-)[24,25]. In the particular case of vibrational spectroscopy ($\omega = \omega_{IR}$), the summation runs over the N vibration modes of a molecule and the amplitudes involve the product of Raman and IR activities of vibration mode i[22].

In the following, we focus on IR-visible SFG and DFG experiments, but the analysis below also applies to the other kinds of resonant second order generation processes. In an actual SFG/DFG experiment, additional effects interfere with the resonant (R) response of the sample. The most obvious one is the presence of a so-called nonresonant (NR) contribution, which appears essentially constant when ω scans the infrared frequency range. For vibrational IR-visible SFG/DFG, the nonresonant signal usually arises from the substrate supporting the molecules. Its amplitude and phase strongly depend on the nature of the substrate[26,27], and on the frequencies and polarization combinations of the light beams. Therefore the general expression for $\chi^{(2)}$ follows:

$$\chi^{(2)} = \chi^{(2)}_{NR} + \chi^{(2)}_{R} = \chi_0 e^{i\Phi} + \sum_{i=1}^{N} \frac{A_i}{\omega - \Omega_i \pm i\Gamma_i} \qquad (4)$$

Additional terms appearing in the calculation of the SFG intensity include Fresnel factors, local field contributions[28], averaging of the molecular responses (β) into the macroscopic $\chi^{(2)}$ [29], surface versus bulk response[30], breaking of the first-level approximations (e.g. higher order multipolar expansion[31]). Most of these effects may be embedded into Equation (4) by using effective susceptibility components.

For the IR-resonant SFG/DFG case, the need for a distinct phase at the numerator of each Lorentzian function may be discussed. In the simple harmonic oscillator description, $A_i$ values are essentially real[22]. However, many authors consider the case where $A_i$ also must have a specific phase. It is possible to list several cases where complex amplitudes may be required:

- embedding of complex Fresnel factors in the effective amplitudes of the resonances[32,33]. As an example, in SFG experiments on an isotropic surface in the ppp polarization combination, the effective $\chi^{(2)}$ is the sum of four components (i.e. zzz, xxz, xzx and zxx)[26]. Considering a vibration mode excited along the z-direction, Fresnel contributions in that case for parallel and perpendicular components of the electric fields contribute differently to xxz and zzz terms. In the case of substrates with complex refractive indices (e.g. a metal or a dielectric near an absorption edge), the phases of Fresnel terms differ;

- the coupling of the SFG process to the localized surface plasmon of nanoparticles[34] or to surface plasmon polaritons in the visible[35] or infrared[36] ranges;

- specific experimental geometries requiring to take into account the local field amplitudes and phases, as is the case in an ATR geometry[37-39];

- embedding of doubly resonant effects in the amplitudes of the resonances[15];

- influence of higher order terms in the multipolar expansion of the source fields of second-order nonlinear processes. Mostly magnetic and quadrupolar contributions have been considered in the literature[19]. For example, resonant quadrupolar terms may be recast in effective second-order susceptibilities comprising Lorentzian terms[40], the phases of which depend on the interference between several contributions involving various quadrupole matrix elements.

- a strong charge transfer between a metallic substrate and adsorbates, whose effect on different vibration modes creates phase shifts between them[41];

- interference between reflected and refracted SF waves in thick structures leading to a sum of $\chi^{(2)}$ components with different phases[42,43];

- additionally, it is not rare to encounter situations for which fitting with real numerators is simply not possible[44-46], even if it is difficult to discriminate between artifactual distortions of the experimental data (due to optical alignments, detection procedures, imperfect calibrations) and theoretical grounds justifying phase shifts between vibrational resonances, as listed above.

From the previous analysis, we deduce that we may write the most general expression for an effective resonant $\chi^{(2)}$ as

$$\chi^{(2)}(\omega) = \chi_0 + \sum_{i=1}^{N} \frac{A_i}{\omega - \omega_i} \qquad (5)$$

where $\chi_0$ is set real and positive, $\omega_i$ stands for $\Omega_i \pm i\Gamma_i$, and phases $\Phi_i$ in equation (3) are all distinct. This function contains the physical chemistry of the sample under study, which lies in the adjustable parameters. However, it is not directly accessible in an experiment, for which only the spectrum $\left|\chi^{(2)}(\omega)\right|^2$ can be measured. In the following, we therefore distinguish the spectrum from its generator $\chi^{(2)}(\omega)$.

Most authors extract the chemical information by numerically fitting the experimental spectrum according to equations (1) and (5) to get the resonance parameters: frequencies, widths and amplitudes. Even if alternate methods exist[47], curve fitting is universally used for IR-resonant SFG experiments, as it provides very fast results and does not require to dig deep into the theoretical concepts underlying the SFG process. As such, it is easily accessible to physical chemists who use SFG as a spectroscopic tool among others.

We point out two drawbacks of numerical fits according to equations (1) and (5):

- various choices of the initial guess for the fit parameters may lead to different final fits. For example, Sgura *et al.*[48] have shown that, in a nonlinear problem for which the best fit is sought by nonlinear least-square adjustment of parameters, it is possible to find several so-called numerical global minima of the objective function, whatever the value of the tolerance used. In other words, a nonlinear fit never ensures that the minimum found is the exact solution sought;

- even in the ideal case of a defect-free algorithm for the fitting procedure, the difference between a generator and its spectrum (i.e. the square modulus operation) raises the question of the singleness of the sets of fit parameters. As an illustration, Le Rille *et al.*[25] have shown that it is possible to find (and calculate) two sets of fit parameters (i.e. two distinct generators) which build exactly the same SFG spectrum made of one vibrational resonance and a non resonant background.

Therefore, it is reasonable that the growing number of users of SHG/SFG/DFG tools question the trust they may put in the results of their fits.

In this article, we show that care must indeed be taken when performing fits of experimental data according to Equations (1) and (5). The uniqueness of the fit parameters is not ensured, which may raise concerns when extracting chemical information from them. We therefore propose a method to calculate all the equivalent sets of fit parameters, which can be used after fitting to improve or even drastically change the fit results, in order to select the most appropriate parameters according to the physical chemistry of the material under study.

As demonstrated in the Appendix, there are, in the general case, for a given number of resonances N, $2^N$ equivalent generators, each described by its own set of parameters $A_i$ (i.e. modulus and phase), generating the same spectrum $I(\omega)$. In the case of vanishing $\chi_0$, there are still $2^{N-1}$ such generators (and associated sets of parameters $A_i$). The Appendix also gives the guidelines for an algorithm calculating the different sets. For all sets of parameters, $\chi_0$ and $\omega_i$ remain the same. This rather surprising result had already been demonstrated by Le Rille et al.[25] for N=1, which is the only case for which an analytical solution can be given using their method. We notice in the examples below that each amplitude $\tilde{A}_i$ distributes in the vicinity of two poles. Combining the ways to choose one of the poles for each coefficient leads to $2^N$ possibilities. This remark does not hold true in the case of a vanishing nonresonant contribution $\chi_0$.

The case where all resonances share a common phase is often used in the literature for the reasons exposed above. It represents a particular case of the more general situation where the dephasings between the resonances are set fixed. In such cases, there is only one free phase parameter (chosen as the phase of resonance 1). In this situations, it is reasonable to wonder whether, among the $2^N$ (resp. $2^{N-1}$) sets of parameters, it is possible to find several ones which fulfil these additional constraints on the phases. The fact that the amplitudes $A_i$ have a fixed phase in the complex plane does not have any direct consequence on the values of roots $J_i$ of the polynomial $J^N$ (cf. Appendix). In fact, the $J_i$'s depend both on $A_i$ and $\omega_i$, so that a particular property of the $A_i$ alone has in general no influence on the generality of the problem. As a consequence, except accidental numerical combinations, there should be only one set of parameters with resonances in phase or having fixed dephasing. However, finding such a function by data fitting is not sufficient to ensure that the best fit has been found, as is illustrated in the second example below.

The most striking consequence of this theoretical analysis is the existence of ghost resonances. Let's consider a generator function built of N resonances (Equation 5) and the associated spectrum (Equation 1). In addition to the $2^N-1$ equivalent generators with N resonances, it is possible to build additional generators having the same spectrum with one (or more) extra resonances, all with nonvanishing amplitudes. The resonances artificially added to the initial set do not show up in the spectrum, whereas they exist as a first order pole in the new generators. This general property defines a ghost resonance. As a consequence, the generator may contain more resonances than the spectrum. In other words, seeing N resonating features on a spectrum only means that there are at least N resonances in the generator. This situation can be related to the case, well-known in SFG spectroscopy, of a nonresonant contribution $\chi_0$ interacting with one resonance of width $\Gamma$, amplitude $2\Gamma\chi_0$ and phase 270° (resp. 90° for DFG), which is mathematically invisible in the SFG signal as the spectrum will appear as a strict flat line. This theoretical case is very close to the experimental spectra from molecules adsorbed on gold and studied by IR-visible SFG in the blue-green region of the visible spectrum[26]. However, as illustrated in the first example below, the ghost resonances also exist without a nonresonant contribution, in which case the invisible resonance interferes destructively with the contributions of its neighbors. The phenomenon of ghost resonances comes from to the fact that, to the existing generator made of N modes with amplitudes chosen among the $2^N$ possibilities, it is possible to add a (N+1)th resonance with a vanishing amplitude and an arbitrary $\omega_{N+1}$. One may then construct in total $2^{N+1}$ equivalent generators. Among these, it is easy to show (cf. Appendix) that only the initial $2^N$ generators have a zero amplitude for the (N+1)th resonance, whereas the $2^N$ others have a nonvanishing amplitude for that resonance.

Turning to the analysis of experimental data, these results raise several issues. We suppose that a generator could be found by the fitting procedure, which accounts for the N resonances showing up on an experimental spectrum. It is then possible to create the $2^N-1$ alternative generators of the same fitting function with N modes. It becomes necessary at that stage to find the one which carries the appropriate chemical information relevant to the case under study. As discussed below, complementary experiments coupled to a fine analysis of fit parameters seem mandatory for that purpose. In a second stage, one must deal with the ghost resonances issue, which may arise in two ways: there may be too many resonances in the generator (i.e. the generator includes ghost resonances) and/or there may be ghost resonances missing in the generator. In the first case, some of the resonances in the generator do not contribute to the spectrum, whereas in the second case the generator misses resonances which should be present even if they do not appear in the spectrum. As the ghost resonances have no effect on the experimental data, it is not possible to discriminate them using only one experimental spectrum. Finally, one must assess whether the ghost resonances found must be kept or discarded on the basis of physical chemistry grounds. Of course, in the case of IR-visible SFG for example, the resonances are usually taken among the IR-active vibration modes of the material. However, the IR and Raman activities may be strongly modified by the adsorption process (e.g. through charge transfer or molecular geometry modification) and the resonances by interactions inside the interface (e.g. by dipole-dipole coupling or Stark shift). These effects make the choice of the number of modes N and the guesses for the ω values of the ghost resonances rather uncertain. As a consequence, there are indeed exactly $2^N$ equivalent sets of parameters to construct a given fit function using exactly N modes, but rather an infinity if one considers an arbitrary nonvanishing number of ghost resonances.

As an illustration of the ghost resonances issue, we take the example of an SFG experiment in which a strong nonresonant background from a gold substrate (either planar or nanoparticles) interferes with $CH_3$ vibration modes from adsorbates (e.g. an alkanethiol). We adapt the following from the example of Figure 6 and Table 2 in Ref 34, which studied dodecanethiol adsorbates on spherical gold nanoparticles. Due to the high signal-to-noise ratio in that experiment, it could have been possible to fit not only with the three $CH_3$ resonances but also with two $CH_2$ modes, as was done for the data of Figure 5. In other words, we can introduce two ghost peaks in the original fit parameters.

We therefore take as original parameters:

$\chi_0 = 1.13$; $\tilde{A}_1 = 1.98$; $\tilde{A}_2 = 1.17$; $\tilde{A}_3 = 2.14$; $\tilde{A}_4 = 0$; $\tilde{A}_5 = 0$;

$\omega_1 = 2881 cm^{-1}$; $\omega_2 = 2943 cm^{-1}$; $\omega_3 = 2967 cm^{-1}$; $\omega_4 = 2853 cm^{-1}$; $\omega_5 = 2917 cm^{-1}$;

$\Gamma_1 = \Gamma_2 = \Gamma_3 = \Gamma_4 = \Gamma_5 = 3.5 cm^{-1}$;

$\Phi_1 = \Phi_2 = \Phi_3 = 214.8°$.

The resonances relate, from 1 to 5, to symmetric $CH_3$ stretching, $CH_3$ Fermi resonance, asymmetric $CH_3$ stretching, $CH_2$ symmetric and $CH_2$ asymmetric stretching modes. The corresponding function is drawn on Figure 1(a). In Table 1, we report, among the 32 sets of parameters, the eight ones with nonvanishing $CH_2$ ghost resonances. The first result is that these ghost resonances have an amplitude around four times greater than the real $CH_3$ ones. It is not obvious from a simple look at the table that they simply do not show up at all in the spectrum. Secondly, we observe the above-mentioned repartition of amplitudes for resonances 1, 2 and 3 around two poles for each, with respective values very far apart, respectively around 3.5, 6.5 and 3 times greater. Among these eight sets, we can define criteria for acceptable sets as is done in Ref 32. As far as the phases are concerned, we expect a common phase for resonances 1, 2 and 3 on one side , and 4 and 5 on the other side, as they relate to the same chemical species. With these criteria, sets 2, 3, 4, 6 and 8 are not acceptable. For sets 5 and 7, the phase span remains acceptable (resp. 35.8°/11.5°; 34.8°/2.8°), whereas it is very good for set 1 (16.5°/14.3°). As a consequence, there is no obvious reason why set 1 could not represent an acceptable fit of experimental data with five resonances. Apart from the knowledge of the existence of the original set, only a critical analysis of the parameters coupled to a study of the experimental curves may lead to the rejection of set 1.

Inversely, the experimental curve could have been fitted with the hypothesis of five vibration modes, and a fit close to set 1 obtained. In that case, applying the algorithm shows that there are eight sets with vanishing, or almost vanishing, amplitudes for modes 4 and 5, hinting at their potential status of ghost resonances. Naturally, the presence of $CH_2$ resonances in the SFG spectrum of an alkanethiol self-assembled film being often used as a sign of gauche defects and disorder in the monolayer[33], the chemical interpretation of the data based on the original set of parameters severely differs from that based on set 1.

The phases of the $CH_2$ resonances are very close to 270° and their amplitudes not far from the quantity $2\Gamma\chi_0$, which explains why they are transparent on the graph. The deviation from this perfect case represents the contribution of the neighboring resonances in the vanishing process. To test its amplitude, we performed the same calculation with a vanishing nonresonant contribution. Results are shown in Table 2, with the corresponding graph in Figure 1(b). We remark in that case that Set 1 is again very close to a realistic solution according to the above-mentioned criteria, as the phase span is very small (16.5°/12.5°). Even if the effect is less spectacular than in the previous case, it is worth noting that the amplitudes of the ghost resonances are not negligible at all, respectively around 0.6 and 0.2 times the amplitude of the smallest real resonance. Contrary to the previous case, the presence of the ghost resonances does not perturb too much the three other ones, both in amplitude and in phase. However, data interpretation is again greatly modified by their existence.

To go further in the use of the model presented here, we take another example from the literature enhancing strong interference between resonant and nonresonant contributions. It is in fact rather difficult to find references where the actual phases of the resonances, when relevant, are listed. It is the case in Ref 44, which presents SFG experimental spectra of a thiophenol monolayer on Ag(111), with three vibration modes of the adsorbates in the 1000cm$^{-1}$ region. Neither the amplitudes and widths of the modes, nor the fit curve are presented in the paper, but it is possible to reconstruct an acceptable function from the known parameters (Figure 2). After a careful comparison between Ref 44 and 49, and with the help of the experimental and fit curves, it is possible to correct an error in Table I of Ref 44, where the phase of the resonance at 1074cm$^{-1}$ should be -95° instead of -5°.

We therefore use for original parameters, with the notations of the present work:

$\chi_0 = 0.45$; $\tilde{A}_1 = 3$; $\tilde{A}_2 = 1$; $\tilde{A}_3 = 1.5$;

$\omega_1 = 1003$cm$^{-1}$; $\omega_2 = 1025$cm$^{-1}$; $\omega_3 = 1074$cm$^{-1}$;

$\Gamma_1 = \Gamma_2 = 2.5$cm$^{-1}$; $\Gamma_3 = 4$cm$^{-1}$;

$\Phi_1 = 55°$ $\Phi_2 = 215°$; $\Phi_3 = 275°$

The nonresonant baseline is not flat, which makes fitting more difficult. The corresponding function is shown on Figure 2, exhibiting a good agreement with Figure 2 of Ref 44. We have calculated the seven other sets of parameters based on this one (Table 3).

The same observations as above about an organization of the amplitudes around two poles with significantly distinct values can be made. In this case, it is also interesting to look closer at sets 2 and 6. For these, the phases of the three resonances are close to the particular case of ideal and undisturbed harmonic vibrations, in which all phases are equal modulo 180°. We therefore used parameters of fit 2 and 6 as initial guesses to try and improve the fits by imposing such a condition on the phases and letting other parameters ($\tilde{A}_i$, $\Gamma_i$) free. Set 6 gave the best results, with the following final parameters:

$\chi_0 = 0.446$; $\tilde{A}_1 = 2.889$; $\tilde{A}_2 = 2.477$; $\tilde{A}_3 = 2.211$;

$\Phi_1 = 74.5°$; $\Phi_2 = \Phi_3 = 254.5°$.

$\Gamma_1 = 2.47$cm$^{-1}$; $\Gamma_2 = 2.79$cm$^{-1}$; $\Gamma_3 = 4.44$cm$^{-1}$.

Set 2 on the contrary led to a bad quality fit for resonance 3. The curve corresponding to this alternate fit issued from set 6, drawn on Figure 2, shows that the quality of the fit is rather good. The amplitudes of modes 2 and 3 exhibit strong variations as compared to the initial parameters. It is therefore possible to find a fit according to Equation (5) with purely real amplitudes. There are indeed discrepancies between the original and the new fit curves, especially for the shape of resonance 3, and we do not claim that the new fit is better than the previous one, in particular because the analysis was not performed on the original data. However, this example illustrates the difficult task of fitting SFG curves *ab initio*, that is without strong conditions imposed on the parameters. It shows that it seems possible to fit the experimental curve with in phase resonances, probably paying the price of a slightly lower quality fit, but within hypotheses which are much easier to justify. As above, the scientific content and conclusions drawn from both the amplitudes and the phases of the three vibration modes will indeed vary much from one fit to the other. Finding such an alternate fit was possible because the method developed here allowed to scan all the equivalent sets of parameters, only a small number of which could lead to satisfactory fitting functions.

As a conclusion, the data presented in this article illustrate the fact that it is not possible to consider a fit of SFG/DFG/SHG data as a unique solution. In the case of out of phase resonances, it is mathematically not true. We propose an algorithm to rapidly find the $2^N$ (resp. $2^{N-1}$ for vanishing nonresonant contribution) alternate sets of parameters building up strictly the same intensity function as a given original set with N resonances. The existence of such alternate sets of parameters leads to the possibility of ghost resonances, which may completely blur the analysis of the experimental data and lead to erroneous conclusions on the structure and dynamics of the interface. Finally, we have illustrated some tricks hidden behind the concept of "best fit". The best fit is always defined within a given frame (fit algorithm, value of the tolerance, initial guesses, signal-to-noise ratio, model describing the process). In particular, considering that there are several local minima of the objective function to the fit problem, the initial guesses play an essential role. Performing a full analysis of all the available alternate sets of parameters may help finding a more accurate and more compatible solution for a given fit, even in the rather classical case where all resonances are set in phase.

Due to the specificity of nonlinear optical spectroscopies, and in particular their high sensitivity to interfacial processes involving very low amounts of matter, it could appear interesting to use and analyze their experimental data in a self-sufficient manner, that is without the help of other more conventional experimental techniques. The results presented here prove that it is most of the times not possible, even after a careful analysis of the fit results. To discriminate between the various sets of parameters, it still seems advisable to establish data analysis on a firm ground with the help of other spectroscopies, for example IR absorption, Raman spectroscopy and HREELS in the case of IR-visible SFG/DFG. They will provide an image, even sometimes distorted as compared to SFG/DFG, of the processes at stake at the molecular level, and help to establish the fits of nonlinear spectroscopy data on a firmer ground. However, this is still not sufficient to prove that the result of the fitting procedure of the second-order nonlinear optical spectroscopies is correct. There is no universal rule to select the appropriate set of parameters among the multiple solutions. Anyway, it is interesting to note that there are some ways to discriminate between the various generators. For example, the comparison between SFG and DFG data[25] greatly helps selecting among several possible sets parameters, and get rid of ghost resonances. The use of phase-sensitive interference methods in nonlinear optics[50] may also give access to the phases of some of the complex quantities in $\chi^{(2)}$. Finally, the evolution of the IR-visible SFG/DFG/SHG spectra with a tunable parameter (e.g. the polarizations of the light beams[51], the electrochemical potential[11] or the visible wavelength[15]) requires to fit series of experimental curves with coherent sets of parameters, usually helping to lift the ambiguity on their values.

**Acknowledgments**

We are greatly indebted to J.C. Léger, Mathematics Department, Université Paris-Sud, Orsay, France for his great help in analyzing the mathematical part of this work. We also warmly thank E.R. Eliel, Leiden Institute of Physics, Netherlands, for providing us with Ref 49.

**Table 1.** Sets of equivalent parameters for curve (a) in Figure 1. The original set is used to build the curve with a vanishing amplitude for resonances 4 and 5, the eight others have nonvanishing amplitudes for these resonances.

|  | original | **Set 1** | Set 2 | Set 3 | Set 4 | Set 5 | Set 6 | Set 7 | Set 8 |
|---|---|---|---|---|---|---|---|---|---|
| $\tilde{A}_1$ | 1.98 | **2.08** | 7.63 | 2.09 | 7.67 | 2.08 | 7.64 | 2.09 | 7.68 |
| $\Phi_1$ (deg.) | 214.8 | **211.8** | 255.1 | 217.1 | 260.5 | 214.8 | 258.1 | 220.2 | 263.5 |
| $\tilde{A}_2$ | 1.17 | **1.22** | 1.22 | 7.98 | 8.02 | 1.24 | 1.25 | 8.17 | 8.21 |
| $\Phi_2$ (deg.) | 214.8 | **195.3** | 190.5 | 244.9 | 240.1 | 205.4 | 200.6 | 255.0 | 250.2 |
| $\tilde{A}_3$ | 2.14 | **2.17** | 2.17 | 2.25 | 2.25 | 6.54 | 6.56 | 6.79 | 6.80 |
| $\Phi_3$ (deg.) | 214.8 | **203.3** | 199.9 | 189.1 | 185.6 | 241.2 | 237.7 | 226.9 | 223.5 |
| $\tilde{A}_4$ | 0 | **8.56** | 8.74 | 8.58 | 8.76 | 8.57 | 8.75 | 8.59 | 8.77 |
| $\Phi_4$ (deg.) | — | **279.0** | 288.8 | 282.7 | 292.6 | 281.3 | 291.1 | 285.0 | 294.8 |
| $\tilde{A}_5$ | 0 | **8.15** | 8.27 | 8.38 | 8.50 | 8.20 | 8.32 | 8.43 | 8.56 |
| $\Phi_5$ (deg.) | — | **264.7** | 256.3 | 277.1 | 268.7 | 269.8 | 261.4 | 282.2 | 273.8 |

**Table 2.** Sets of equivalent parameters for curve (b) in Figure 1. The original set is used to build the curve with a vanishing amplitude for resonances 4 and 5, the four others have nonvanishing amplitudes for these resonances.

|  | Original | **Set 1** | Set 2 | Set 3 | Set 4 |
|---|---|---|---|---|---|
| $\tilde{A}_1$ | 1.98 | **2.08** | 2.14 | 2.09 | 2.15 |
| $\Phi_1$ (deg.) | 0 | **0** | 0 | 0 | 0 |
| $\tilde{A}_2$ | 1.17 | **1.22** | 1.24 | 1.50 | 1.53 |
| $\Phi_2$ (deg.) | 0 | **-16.5** | -42.2 | 13.9 | -11.8 |
| $\tilde{A}_3$ | 2.14 | **2.17** | 2.18 | 2.41 | 2.43 |
| $\Phi_3$ (deg.) | 0 | **-8.4** | -29.4 | -40.0 | -61.0 |
| $\tilde{A}_4$ | 0 | **0.72** | 0.73 | 0.72 | 0.73 |
| Φ$_4$ (deg.) | — | **99.3** | 92.3 | 97.7 | 90.8 |
| Ã$_5$ | 0 | **0.23** | 0.31 | 0.24 | 0.31 |
| Φ$_5$ (deg.) | — | **86.8** | 32.0 | 92.3 | 37.5 |

**Table 3.** The eight sets of equivalent parameters for the original curve in Figure 3.

| | Original | Set 1 | **Set 2** | Set 3 | Set 4 | Set 5 | **Set 6** | Set 7 |
|---|---|---|---|---|---|---|---|---|
| Ã$_1$ | 3 | 5.11 | **3.05** | 5.20 | 3.00 | 5.12 | **3.05** | 5.20 |
| Φ$_1$ (deg.) | 55 | 296.0 | **63.8** | 304.8 | 56.2 | 297.2 | **65.0** | 306.0 |
| Ã$_2$ | 1 | 1.05 | **2.52** | 2.65 | 1.00 | 1.05 | **2.52** | 2.65 |
| Φ$_2$ (deg.) | 215 | 183.8 | **262.5** | 231.3 | 216.7 | 185.5 | **264.2** | 233.0 |
| Ã$_3$ | 1.5 | 1.52 | **1.51** | 1.53 | 2.17 | 2.19 | **2.18** | 2.21 |
| Φ$_3$ (deg.) | 275 | 263.9 | **270.6** | 259.5 | 274.3 | 263.3 | **269.9** | 258.8 |

Figure captions

**Figure 1**. (a) Simulated SFG spectrum made of three $CH_3$ resonances and a nonresonant contribution, according to parameters described in text (squares). (b) Simulated SFG spectrum made of the same $CH_3$ resonances without nonresonant contribution (circles). Continuous curves are the same mathematical functions, drawn using sets 1 in Tables 1 and 2 respectively, and comprising five resonances.

**Figure 2**. Simulated SFG spectra based on the experimental curves and the fit parameters of Ref 43. The black curve (squares) is the closest to the experimental data. The grey curves (circles) was created from an alternate set of fit parameters by imposing the three resonances in phase.

Figures

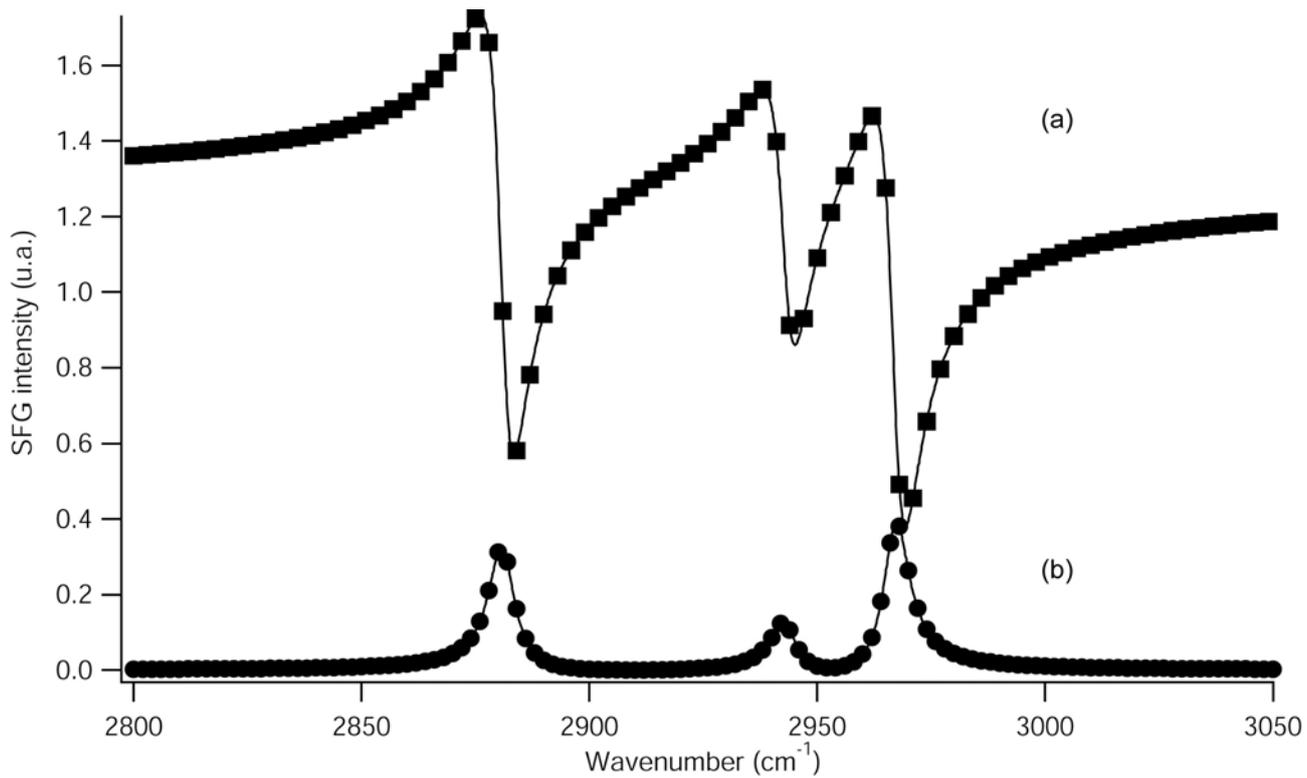

Figure 1

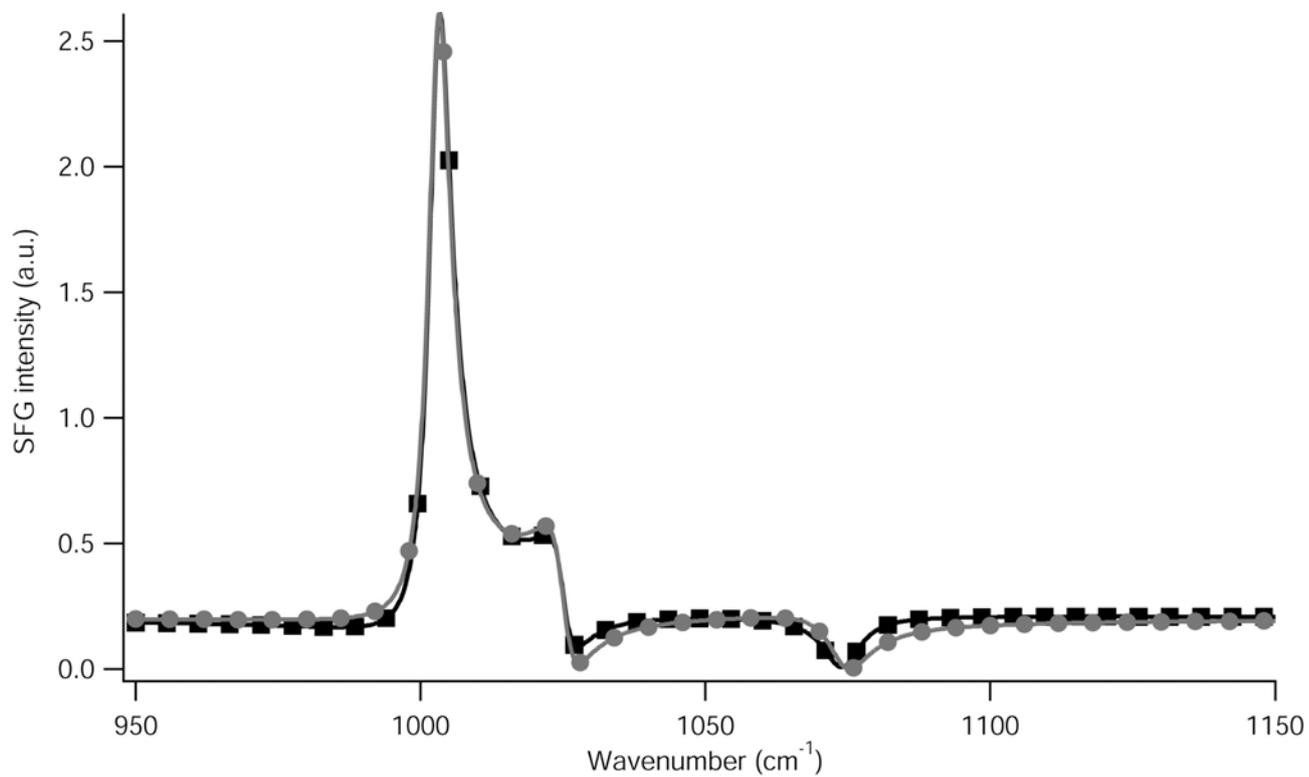

Figure 2

**Appendix**

Starting from Eq. (5) and the notations of the text, we write the generator:

$$\chi^{(2)} = \chi_0 + \sum_{i=1}^{N} \frac{A_i}{\omega - \omega_i}$$ with $\chi_0$ real and positive. All $\omega_i$ are supposed distinct one from the others.

The 2N+1 adjustable parameters fitting the experimental spectrum are: $\chi_0$, $A_i$, $\omega_i$. The question is: knowing one set of parameters, are there other sets (i.e. other generators) building the same spectrum, and how many ? We postulate the existence of such another set of 2N'+1 parameters $\chi'_0$, A'i, ω'i. In other words, we define two functions of the real variable ω

$$J(\omega) = \chi_0 + \sum_{i=1}^{N} \frac{A_i}{\omega - \omega_i}$$ for which the values of $\chi_0 \geq 0$, $A_i$, $\omega_i$ are known, and

$$K(\omega) = \chi'_0 + \sum_{i=1}^{N'} \frac{A'_i}{\omega - \omega'_i}$$, such as $|J(\omega)|^2 = |K(\omega)|^2$ for any positive real number ω.

Considering the spectrum as a complex rational fraction with respect to variable ω, its partial fraction decomposition leads to

$$|J(\omega)|^2 = \chi_0^2 + \sum_{i=1}^{N} A_i \left( \chi_0 + \sum_{j=1}^{N} \frac{\overline{A_j}}{\omega_i - \overline{\omega_j}} \right) \frac{1}{\omega - \omega_i} + \sum_{i=1}^{N} \overline{A_i} \left( \chi_0 + \sum_{j=1}^{N} \frac{A_j}{\overline{\omega_i} - \omega_j} \right) \frac{1}{\omega - \overline{\omega_i}} \quad (A1)$$

where the bars stand for complex conjugation. The last two sums in (A1) are complex conjugates of one another for a real ω. Parameters $\omega_i$ having a fixed sign for their nonvanishing imaginary parts, they cannot be confused with their conjugates.

We define at this stage a ghost resonance i as one for which the coefficient of pole $\omega_i$ in (A1) vanishes. This implies that it is not visible in the spectrum. We first analyze the problem in the absence of ghost resonances in J and K.

The equality of the spectra $|J(\omega)|^2 = |K(\omega)|^2$ must hold for any positive real number ω. The properties of the decomposition of complex rational fractions imply that

- $\chi'^2_0 = \chi_0^2$ (uniqueness of quotient);

- the unprimed resonances correspond one to one to the primed ones (uniqueness of the partial fractions). This is equivalent to the fact that N equals N', and that a renumbering exists such that $\omega'_i = \omega_i$ (i.e. $\Omega'_i = \Omega_i$ and $\Gamma'_i = \Gamma_i$).

Only $A'_i$ may thus differ from $A_i$ from one set of parameters to the other. The problem is therefore reduced as follows: considering

$$J(\omega) = \chi_0 + \sum_{i=1}^{N} \frac{A_i}{\omega - \omega_i}; \quad K(\omega) = \chi_0 + \sum_{i=1}^{N} \frac{A'_i}{\omega - \omega_i}, \text{ and } |J(\omega)|^2 = |K(\omega)|^2$$

for any positive real number $\omega$, what are the values of $A'_i$ and how many possibilities are there to build such a K function?

1) We first assume that $\chi_0 \neq 0$.

Writing $J(\omega) = \frac{\chi_0 J^N(\omega)}{D(\omega)}$ and $K(\omega) = \frac{\chi_0 K^N(\omega)}{D(\omega)}$ with $D(\omega) = \prod_{i=1}^{N}(\omega - \omega_i)$, $J^N$ and $K^N$ are monic polynomial functions of degree N, they have exactly N complex roots $J_i$ and $K_i$, respectively

$$J^N(\omega) = \prod_{i=1}^{N}(\omega - J_i) \text{ and } K^N(\omega) = \prod_{i=1}^{N}(\omega - K_i).$$

Let $P(\omega)$ be the greatest common divider of $J^N(\omega)$ and $K^N(\omega)$, so that $J^N(\omega) = P(\omega) A(\omega)$ and $K^N(\omega) = P(\omega) B(\omega)$.

Calling $Q = \frac{J(\omega)}{K(\omega)} = \frac{J^N(\omega)}{K^N(\omega)} = \frac{A(\omega)}{B(\omega)}$ we have $|Q|^2 = 1$ for $\omega$ real and Q belongs to the unit circle of the complex plane.

The transformation $z = \frac{\omega - i}{\omega + i}$ maps the real axis to the unit circle, its reciprocal is $\omega = i\frac{1+z}{1-z}$. Therefore Q, as a function of z, is a rational fraction which maps the unit circle onto itself. As such, the roots of Q are exactly the images of its poles under the transformation $z \to \frac{1}{\bar{z}}$. The reader may refer to Ref. 52 for more details.

For α$_i$ a root of Q, let us introduce the Blaschke factors B$_i$ [53], defined by $B_i = \dfrac{z - \alpha_i}{1 - \overline{\alpha_i} z}$, it is clear that $Q = \zeta \prod_{\alpha_i \in R} B_i$, with ζ constant, $|\zeta| = 1$, where the product runs over R, the roots of Q.

Turning back to the ω variable gives $Q = \zeta' \prod_{\beta_i \in R'} B_i, B_i = \dfrac{\omega - \beta_i}{\omega - \overline{\beta_i}}, \beta_i = i\dfrac{1 + \alpha_i}{1 - \alpha_i}$.

The limit of Q as ω → ∞ shows that ζ' = 1.

Therefore $Q(\omega) = \prod_{\beta_i \in R'} \dfrac{\omega - \beta_i}{\omega - \overline{\beta_i}}$, which means that $\beta_i$ belong to the roots of A(ω) (i.e. to the J$_i$'s) and $\overline{\beta_i}$ to that of B(ω).

The monic polynomials A(ω) and B(ω) having the same degree and no common roots, it implies that $\beta_i$ and $\overline{\beta_i}$ describe exactly the roots of A and B, respectively. In other words, $J_i \equiv \beta_i$ and $K_i \equiv \overline{\beta_i}$, and B(ω) is the complex conjugate of A(ω) for real values of ω.

Knowing the N roots J$_i$ of polynomial J$^N$, the N roots of K$^N$ are built by choosing N-m roots among the J$_i$'s while the last m roots are the conjugates of the remaining J$_i$'s, with m taking any value between 1 and N.

$$K^N(\omega) = \prod_{i=1}^{N-m}(\omega - J_i) \prod_{i=N-m}^{m}\left(\omega - \overline{J_i}\right), 1 \leq m \leq N.$$

The number of different possibilities to build a polynomial K$^N$ reduces to the number of partitions into two sets of the set made of the N roots of J$^N$. In the general case, for which all roots are distinct and not real, there are exactly 2$^N$ ways to build K$^N$ polynomials. Consequently, the general number of different sets of parameters A$_i$ to build a unique spectrum is also 2$^N$.

As mentioned in the text, the case N=1, with two different sets of solutions, has been evidenced by Le Rille *et al* [25]. Using the present algorithm also makes it possible to analytically calculate the case N=2. For N>2, a numerical evaluation is necessary.

The reasoning presented here additionally gives an algorithm for such a numerical calculation of the $2^N$ sets of parameters once one of them is known, working with a fixed set of resonances $\omega_i$, following the successive steps:

- knowing J($\omega$), reduce it to common denominator D($\omega$);

- extract numerator and divide it by $\chi_0$ to get $J^N(\omega)$;

- numerically solve the polynomial $J^N$ to find the N roots;

- create the $2^N$-1 $K^N$ polynomials

- for each $K^N$ polynomial, the complex $A'_i$ are evaluated by partial fraction decomposition of

$K(\omega) = \dfrac{\chi_0 K^N(\omega)}{D(\omega)}$ which leads to

$$A'_i = \dfrac{\chi_0 K^N(\omega_i)}{\prod_{\substack{j=1 \\ j \neq i}}^{N}(\omega_i - \omega_j)} \qquad (A2)$$

2) The case $\chi_0 = 0$ is treated in the same way, apart from a few differences. For $J(\omega) = \sum_{i=1}^{N} \dfrac{A_i}{\omega - \omega_i}$ there is no absolute phase reference (contrary to the previous section, in which setting $\chi_0$ real and positive fixes the phases). As a consequence, the $A_i$ are only determined modulo a phase shift, which would result in the same $\left|\chi^{(2)}\right|^2 = \left|J(\omega)\right|^2$. To fix the phase in a symmetrical way, we impose $\sum_{i=1}^{N} A_i$ to be real and positive.

Then J transforms into $J(\omega) = \left(\sum_{i=1}^{N} A_i\right) \dfrac{J^{N-1}(\omega)}{D(\omega)}$ where $J^{N-1}$ is a monic polynomial of order N-1. Assuming that $\sum_{i=1}^{N} A_i$ does not vanish, the problem is solved in the same way as in the previous case, $\sum_{i=1}^{N} A_i$ playing the role of $\chi_0$ and $J^{N-1}$ that of $J^N$.

We define as above $K(\omega) = \sum_{i=1}^{N} \frac{A'_i}{\omega - \omega_i} = \left(\sum_{i=1}^{N} A'_i\right) \frac{K^{N-1}(\omega)}{D(\omega)}$ with the condition $|J(\omega)|^2 = |K(\omega)|^2$.

The asymptotic behavior of this equality for $\omega \to +\infty$ implies that $\left|\sum_{i=1}^{N} A_i\right|^2 = \left|\sum_{i=1}^{N} A'_i\right|^2$. Choosing $\sum_{i=1}^{N} A'_i$ real and positive (which fixes the phase for K) leads to $\sum_{i=1}^{N} A_i = \sum_{i=1}^{N} A'_i$. The analysis over the N-1 roots of $K^{N-1}(\omega)$ is led in the same way as above for $K^N(\omega)$, leading to the conclusion that there are $2^{N-1}$ equivalent sets of parameters in that case.

If $\sum_{i=1}^{N} A_i$ vanishes, the analysis is performed using the same scheme. In a general way, $J(\omega)$ may be written as the division of a polynomial by $D(\omega)$. One selects the nonvanishing $\omega^p$ term of the numerator with highest order p (p $\leq$ N) to create the $J^p(\omega)$ monic polynomial as above. Repeating the previous analysis shows that the number of equivalent sets of parameters becomes $2^p$.

3) Both cases above have been analyzed in the absence of ghost resonances. Starting from a generator $J(\omega)$ with N poles and free from ghost resonances, it is easy to build generators with N+1 poles having the same spectrum. We add a (N+1)th term to the initial generator, with $A_{N+1} = 0$. For this particular case, the value of $\omega_{N+1}$ is arbitrary, but it will naturally be chosen among meaningful values deduced from the chemical physics properties of the sample under study.

Applying the algorithm above to this new generator, we create a total amount of $2^{N+1}$ generators (resp. $2^N$ if $\chi_0 = 0$). The condition $A_{N+1} = 0$ implies that the complex quantity $\omega_{N+1}$ belongs to the roots of polynomial $J^{N+1}$. Considering the way polynomials $K^{N+1}$ are built, half of them also has $\omega_{N+1}$ as a root, the other half having $\overline{\omega_{N+1}}$, and therefore not $\omega_{N+1}$, as the corresponding root. Consequently, from Eq. (A2) it follows that one half of the generators (labeled $N_1$) have $A_{N+1} = 0$, but for the other half $A_{N+1} \neq 0$ (labeled $N_2$). For these latter, the resonance N+1 has a nonvanishing amplitude, even if it does not contribute to the spectrum, it is thus a ghost resonance. It must be stressed that the value of amplitude $A_{N+1}$ depends on the choice of the parameter $\omega_{N+1}$.

For generators in $N_2$, we therefore have $\overline{K(\omega_{N+1})} = \chi_0 + \sum_{j=1}^{N+1} \frac{\overline{A'_j}}{\overline{\omega_{N+1} - \omega_j}} = 0$, which in turn proves that pole $\omega_{N+1}$ disappears indeed from (A1), and allows to check the status of ghost resonance. We understand that there are in fact two ways for a resonance $\omega_i$ not to appear in a spectrum (A1): as a consequence of either $A_i = 0$ or $\left(\chi_0 + \sum_{j=1}^{N} \frac{\overline{A_j}}{\overline{\omega_i - \omega_j}}\right) = 0$. In the former trivial case, the resonance does not belong to the poles of the generator; in the latter situation, resonance $\omega_i$ is a true ghost resonance according to the definition given in the article. We can thus formulate a more accurate definition of a ghost resonance: for a given generator $J(\omega) = \chi_0 + \sum_{i=1}^{N} \frac{A_i}{\omega - \omega_i}$, resonance $\omega_i$ is a ghost if $J(\overline{\omega_i}) = 0$.

Table of Content Graphic

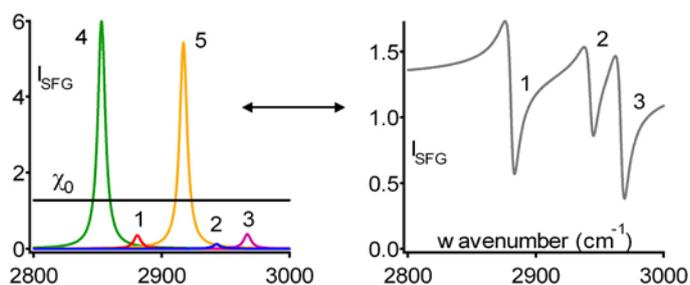

Table of Content Caption

A typical case of ghost resonances in SFG spectroscopy. Vibration modes 4 and 5 ($CH_2$ vibrations) belong to the nonlinear susceptibility whereas they disappear in the SFG spectrum. Left panel shows the five resonances (three $CH_3$ and two $CH_2$ modes) plotted separately and the non-resonant background (from a gold substrate). Right panel displays the corresponding SFG spectrum.